\documentclass{article}

\usepackage[preprint]{neurips_2023}

\usepackage{amsmath,amsfonts,bm}

\def\Figref#1{Fig.~\ref{#1}}

\def\Secref#1{Section~\ref{#1}}

\def\eqref#1{equation~\ref{#1}}
\def\Eqref#1{Equation~\ref{#1}}

\def\1{\bm{1}}

\def\ve{{\bm{e}}}

\def\vh{{\bm{h}}}

\def\vm{{\bm{m}}}

\def\vo{{\bm{o}}}

\def\vw{{\bm{w}}}
\def\vx{{\bm{x}}}

\def\mH{{\bm{H}}}

\DeclareMathAlphabet{\mathsfit}{\encodingdefault}{\sfdefault}{m}{sl}
\SetMathAlphabet{\mathsfit}{bold}{\encodingdefault}{\sfdefault}{bx}{n}

\usepackage{natbib}
\setcitestyle{numbers,square}
\usepackage[utf8]{inputenc} 
\usepackage[T1]{fontenc}    
\usepackage{hyperref}       
\usepackage{url}            
\usepackage{booktabs}       
\usepackage{amsfonts}       
\usepackage{nicefrac}       
\usepackage{microtype}      
\usepackage{xcolor}         
\usepackage{graphicx}
\usepackage{multirow}
\usepackage{diagbox}
\usepackage{enumitem}
\usepackage{threeparttable}
\usepackage{amsthm}
\newtheorem{definition}{Definition}
\usepackage[ruled,linesnumbered]{algorithm2e}
\usepackage{wrapfig}

\AtBeginDocument{%
  \providecommand\BibTeX{{%
    \normalfont B\kern-0.5em{\scshape i\kern-0.25em b}\kern-0.8em\TeX}}}

\newcommand{\modelname}{MBP\xspace}

\newcommand{\datasetname}{ChEMBL-Dock\xspace}
\newcommand{\vpara}[1]{\noindent\textbf{#1 }}

\newcommand{\beq}[1]{{\small\begin{equation}#1\end{equation}}}

\newcommand{\besp}[1]{\begin{split}#1\end{split}}

\newcommand{\norm}[2]{\left\lVert#1\right\rVert_{#2}}
\newcommand{\abs}[1]{\left\lvert#1\right\rvert}

\begin{document}

\title{
Multi-task Bioassay Pre-training for Protein-ligand Binding Affinity Prediction
}
\author{
Jiaxian Yan$^{1,2 *}$, Zhaofeng Ye$^{2}$\thanks{These authors contribute equally to this work.} \,, Ziyi Yang$^2$, Chengqiang Lu$^1$, \\ 
\textbf{Shengyu Zhang$^2$, Qi Liu$^{1\dag}$, Jiezhong Qiu$^{2}$\thanks{Corresponding Authors.} }\\
$^1$University of Science and Technology of China, $^2$Tencent Quantum Laboratory \\
$^\dag$qiliuql@ustc.edu.cn;  jiezhongqiu@outlook.com
}

\maketitle
\vspace{-0.3in}
\begin{abstract}

Protein-ligand binding affinity (PLBA) prediction is the fundamental task in drug discovery. Recently, various deep learning-based models predict binding affinity by incorporating the three-dimensional structure of protein-ligand complexes as input and achieving astounding progress. However, due to the scarcity of high-quality training data, the generalization ability of current models is still limited. In addition, different bioassays use varying affinity measurement labels (i.e., IC50, Ki, Kd), and different experimental conditions inevitably introduce systematic noise, which poses a significant challenge to constructing high-precision affinity prediction models. 
To address these issues, we (1) propose \textit{\textbf{M}ulti-task \textbf{B}ioassay \textbf{P}re-training} (\modelname), a pre-training framework for structure-based PLBA prediction;
(2) construct a pre-training dataset called \datasetname with more than 300k experimentally measured affinity labels and about 2.8M docked three-dimensional structures.
By introducing multi-task pre-training to treat the prediction of different affinity labels as different tasks and classifying relative rankings between samples from the same bioassay, MBP learns robust and transferrable structural knowledge from our new \datasetname dataset with varied and noisy labels.
Experiments substantiate the capability of MBP as a general framework that can improve and be tailored to mainstream structure-based PLBA prediction tasks. To the best of our knowledge, MBP is the first affinity pre-training model and shows great potential for future development. 
All codes and data are available on the online platform \url{https://github.com/jiaxianyan/MBP}.

\end{abstract}

\vspace{-0.2in}

\section{Introduction}
\label{sec:intro}
Protein-ligand binding affinity (PLBA) is a measurement of the strength of the interaction between a target protein and a ligand drug \cite{RIZZUTI2020309}. Accurate and efficient PLBA prediction is the central task for the discovery and design of effective drug molecules \textit{in silico} \cite{Seo2021BindingAP}. 
Traditional computer-aided drug discovery tools use scoring functions (SF) to estimate PLBA roughly\cite{Jacob2008ProteinligandIP}, which is of low accuracy. Molecular dynamics simulation methods can achieve more accurate binding energy estimation \cite{deng2009computations}, but these methods are typically expensive in terms of computational resources and time. In recent years, deep learning (DL) models have been widely used to predict PLBA, which are thought to be promising tools for accurately and rapidly predicting PLBA. Based on the accumulated biological data, a series of DL-based scoring functions have been built, such as Pafnucy \cite{Pafnucy2018}, OnionNet \cite{Zheng2019OnionNetAM}, Transformer-CPI \cite{Chen2020TransformerCPIIC}, IGN \cite{jiang2021interactiongraphnet}, and SIGN \cite{Li2021StructureawareIG}.
In particular, structure-based DL models that use the 3D structure of protein-ligand complexes as inputs are most successful, which typically use 3D convolutional neural networks (3D-CNNs)\cite{Jimnez2018KDEEPPA, HassanHarrirou2020RosENetIB, Jones2020ImprovedPB} or graph neural networks (GNNs) \cite{jiang2021interactiongraphnet, Li2021StructureawareIG} to model and extract the interactions within the protein-ligand complex structures. However, the generalizability of these data-driven DL models is limited because the number of high-quality samples in PDBbind used for model training is relatively small (approximately 5,000) \cite{Liu2017ForgingTB}.

 \begin{wrapfigure}{r}{0.6\textwidth}
  \centering
  \includegraphics[width=\linewidth]{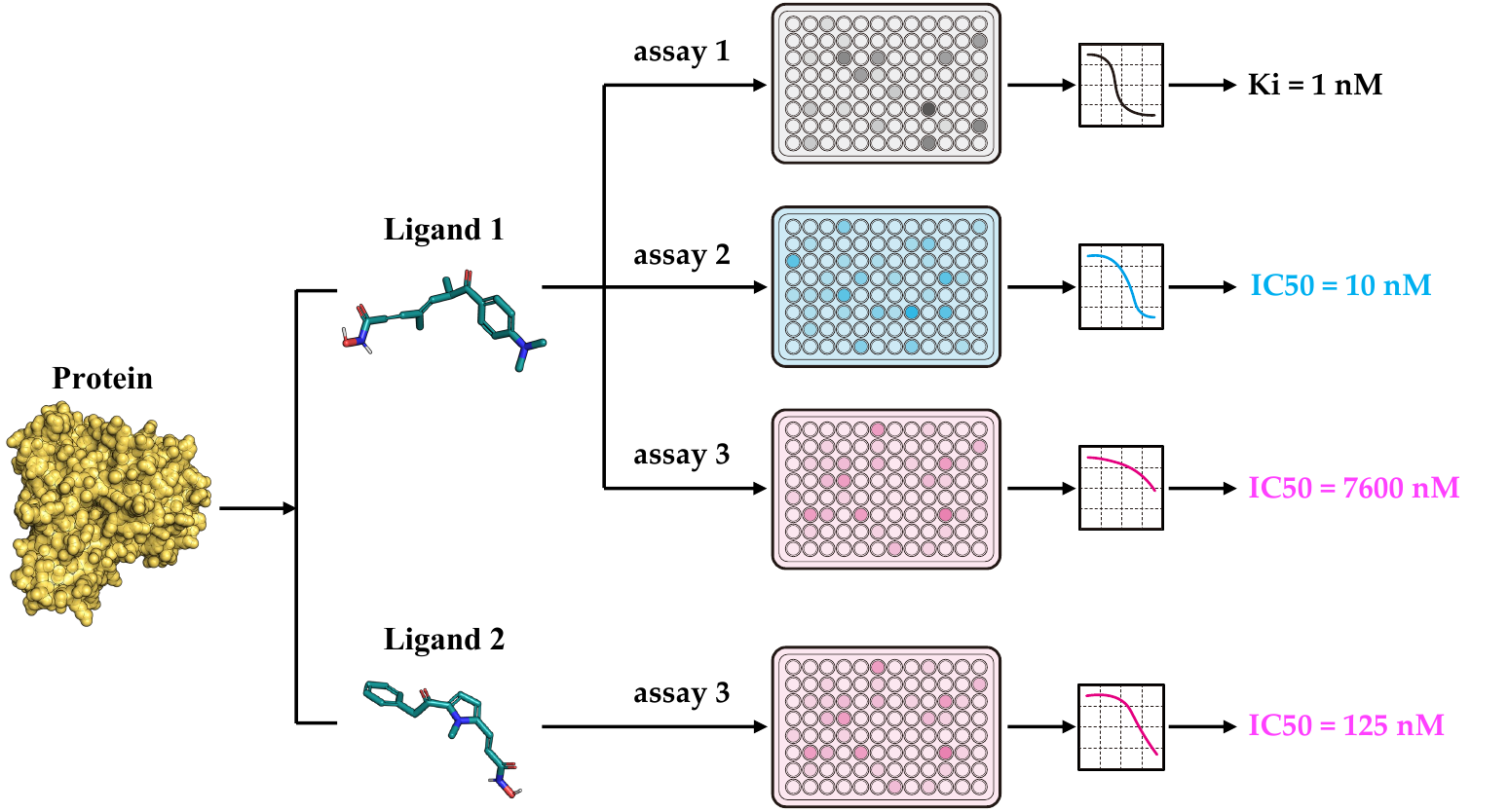}
  \caption{A real example of bioassay data in ChEMBL.
  (1) The top three panels show an example where the same protein-ligand pair have different binding affinities with assay 1-3 in terms of measurement type (IC50 v.s. Ki) and value (IC50=10~nM v.s. IC50=7600~nM). (2) The bottom two panels show an example of the binding of different ligands (Ligand 1 \& 2) to a protein in the same assay (assay 3).}
  \label{fig:1_assay}
\end{wrapfigure}

One solution for this problem is pre-training, which has been widely used in computational biology, such as molecular pre-training for compound property prediction \cite{Zhang2021MotifbasedGS, Maziarka2020MoleculeAT, NEURIPS2020_94aef384, Zhu2022Unified2A} and protein pre-training for protein folding \cite{Fang2021ChemRLGEMGE, Unsal2022LearningFP, pmlr-v139-rao21a, Elnaggar2020ProtTransTC}. These pre-training models utilize data from large-scale datasets to learn embeddings, which expand the ligand chemical space and protein diversities. Therefore, affinity pre-training models on the large amount of affinity data in databases such as ChEMBL \cite{Gaulton2011ChEMBLAL} and BindingDB \cite{Liu2006BindingDBAW} can be helpful. Nevertheless, though attempts have been made to directly use the data, such as BatchDTA \cite{luo2022batchdta}, several challenges have prevented researchers from widely using ChEMBL data for PLBA previously. Firstly, the data were collected from various bioassays, which introduce different system biases and noises to the data and make it difficult for comparison \cite{luo2022batchdta, papadatos2015activity} (\textbf{label noise problem}). For some cases, the affinities of the same protein-ligand pair in different bioassay can have a difference of several orders of magnitude (Fig. \ref{fig:1_assay}). Secondly, several types of affinity measurement exist (\textbf{label variety problem}), such as half-maximal inhibitory concentration (IC50), inhibition constant (Ki), inhibition ratio, dissociation constant (Kd), half-maximal effective concentration (EC50), etc., which cannot be compared directly as well. Thirdly, the unavailability of 3D structures of protein-ligand complexes within ChEMBL poses a significant limitation for researchers in training and leveraging structure-based DL models (\textbf{missing conformation problem}).

To solve the above problems, we propose the Multi-task Bioassay Pre-training (\modelname) framework for structure-based PLBA prediction models. In general, by introducing multi-task and pairwise ranking within bioassay samples, \modelname can make use of the noisy data in databases like ChEMBL. Specifically, the multi-task learning strategy \cite{Crawshaw2020MultiTaskLW} treats the prediction of different label measurement types (IC50/Ki/Kd) as different tasks, thus enabling information extraction from related but different affinity measurements.
Meanwhile, although different assays can introduce different types of noises to the data, data from the same assay is relatively more comparable.
Inspired by recent progress in the recommendation system \cite{Wang2022MP2AM, Cinar2020AdaptivePL, Lei2017AlternatingPL}, by considering ranking between samples from the same assay, the model is enforced to learn the relative relationship of samples and differences in protein-ligand interactions, which allows the MBP to learn robust and transferrable structural knowledge beyond the noisy labels.

We then construct a pre-training dataset, \datasetname, for \modelname.
\datasetname contains 313,224 protein-ligand pairs with from 21,686 assays and the corresponding experimental PLBA labels (IC50/Ki/Kd). Molecular docking softwares are emplyed to generate about 2.8M docked 3D complex structures in \datasetname. Then we implant MBP with simple and commonly used GNN models, such as GCN \cite{kipf2017semisupervised}, GIN \cite{xu2018how}, GAT \cite{veličković2018graph}, EGNN \cite{Han2022GeometricallyEG}, and AttentiveFP \cite{Xiong2020PushingTB}. Experiments on the PDBbind core set and the CSAR-HiQ dataset have shown that even simple models can be improved and achieve comparable or better performances than the state-of-the-art (SOTA) models with MBP. 
Through ablation studies, we further validate the importance of multi-task strategy and bioassay-specific ranking in \modelname.

Overall, the contributions of this paper can be summarized as follows:
\begin{itemize}[leftmargin=*]
\vspace{-0.05in}
    \item We propose the first PLBA pre-training framework \modelname, which can significantly improve the accuracy and generalizability of PLBA prediction models.
\vspace{-0.05in}
    \item We construct a high-quality pre-training dataset, \datasetname, based on ChEMBL, which significantly enlarges existing PLBA datasets in terms of chemical diversities.
\vspace{-0.05in}
    \item We show that even vanilla GNNs can significantly outperform the previous SOTA method by following the pre-training protocol in \modelname.
\end{itemize}

 \vspace{-0.05in}
\section{Related Work}
\label{sec:related}
 \vspace{-0.05in}
\vpara{Protein-Ligand Binding Affinity Prediction.}
One critical step in drug discovery is scoring and ranking the predicted protein-ligand binding affinity. Scoring functions can be roughly divided into four main types: force-field-based, empirical-based, machine-learning-based, and 3D-structure-based \cite{Li2021StructureawareIG}. Force-field-based methods aim at estimating the free energy of the binding by using the first principles of statistical mechanics \cite{Zheng2019OnionNetAM}. Despite its remarkable performance as a gold standard, it suffers from high computational overhead.  Empirical-based methods \cite{gohlke2000knowledge, trott2010autodock, wang2002further} design docking and scoring functions especially to make affinity predictions, while expert domain knowledge is needed to encode internal biochemical interactions.
Machine-learning-based methods, such as random forest \cite{ballester2010machine} and support vector machines (SVM)\cite{kinnings2011machine}, aim to predict binding affinity based on a data-driven learning paradigm. 
These methods, however, rely on the quality of hand-crafted features and have poor performance in generalization. Recently, due to advances in deep learning methods and the creation of structure-based protein-ligand complex datasets, many structure-based deep learning methods \cite{StepniewskaDziubinska2018DevelopmentAE, Zheng2019OnionNetAM, nguyen2021graphdta, danel2020spatial,Maziarka2020MoleculeAT, Gasteiger2020Directional, Song2020CommunicativeRL, jiang2021interactiongraphnet, Li2021StructureawareIG} have been developed for predicting binding affinity. Such methods directly learn the structural information of protein-ligand complexes end-to-end, avoiding artificial feature design. However, due to the scarcity of high-quality training data, current methods still suffer from poor generalization in real applications.

\vpara{Datasets of Protein-Ligand Binding Affinity.}
Existing protein-ligand binding affinity datasets can be roughly divided into three categories. The first category includes datasets such as PDBbind, BindingMOAD \cite{Kramer2010LeaveClusterOutCI}, and CSAR-HIQ \cite{Dunbar2013CSARDS}, which contain 3D co-crystal structures of protein-ligand complexes determined by structural characterization methods and experimentally determined binding affinity values. Such datasets have small yet high-quality data and are typically widely used for training structure-based deep learning models \cite{lu2022tankbind, Strk2022EquiBindGD}. The second category contains large-scale protein-ligand binding affinity measured labels but without 3D structures, such as ChEMBL and BindingDB.
The third category contains the 3D structure and binding affinity value of the protein-ligand complex calculated by molecular docking \cite{Meng2011MolecularDA}, and the representative database is CrossDocked \cite{Francoeur2020ThreeDimensionalCN}. Due to the lack of experimental affinity labels, such datasets are often used to train generative models rather than affinity prediction models  \cite{Peng2022Pocket2MolEM}.

\vpara{Pre-training for Biomolecules.}
Much effort has been devoted to biomolecular pre-training to achieve better performance on related tasks. 
For small molecules and proteins, a series of self-supervised pre-training methods based on molecular graphs \cite{Zhang2021MotifbasedGS, Maziarka2020MoleculeAT, NEURIPS2020_94aef384, Zhu2022Unified2A} and protein sequences \cite{Fang2021ChemRLGEMGE, Unsal2022LearningFP, pmlr-v139-rao21a, Elnaggar2020ProtTransTC} have been proposed, respectively.
However, these existing pre-training methods are designed for individual molecules \cite{zhou2022uni}, and there is still a gap in the research on pre-training methods for protein-ligand affinity.


\vpara{Pairwise Learning to Rank.}
 Learning to Rank (LTR) is an essential research topic in many areas, such as information retrieval and recommendation systems \cite{Cao2007LearningTR, 10.1007/978-3-030-46133-1_15, Liu2009LearningTR}. The common solutions of LTR could be basically categorized into three types: pointwise, pairwise, and listwise. Among these methods, pairwise LTR models are widely used in practice due to their efficiency and effectiveness.
 These years have witnessed the success of pairwise methods, such as BPR \cite{rendle2009bpr}, RankNet \cite{burges2005learning}, GBRank \cite{zheng2007regression}, and RankSVM \cite{lee2014large}.
 In addition, recent studies have shown that the bias between labels can be effectively solved using pairwise methods \cite{Wang2022MP2AM}.




   \begin{figure*}[t]
  \centering
  \includegraphics[width=0.89\linewidth]{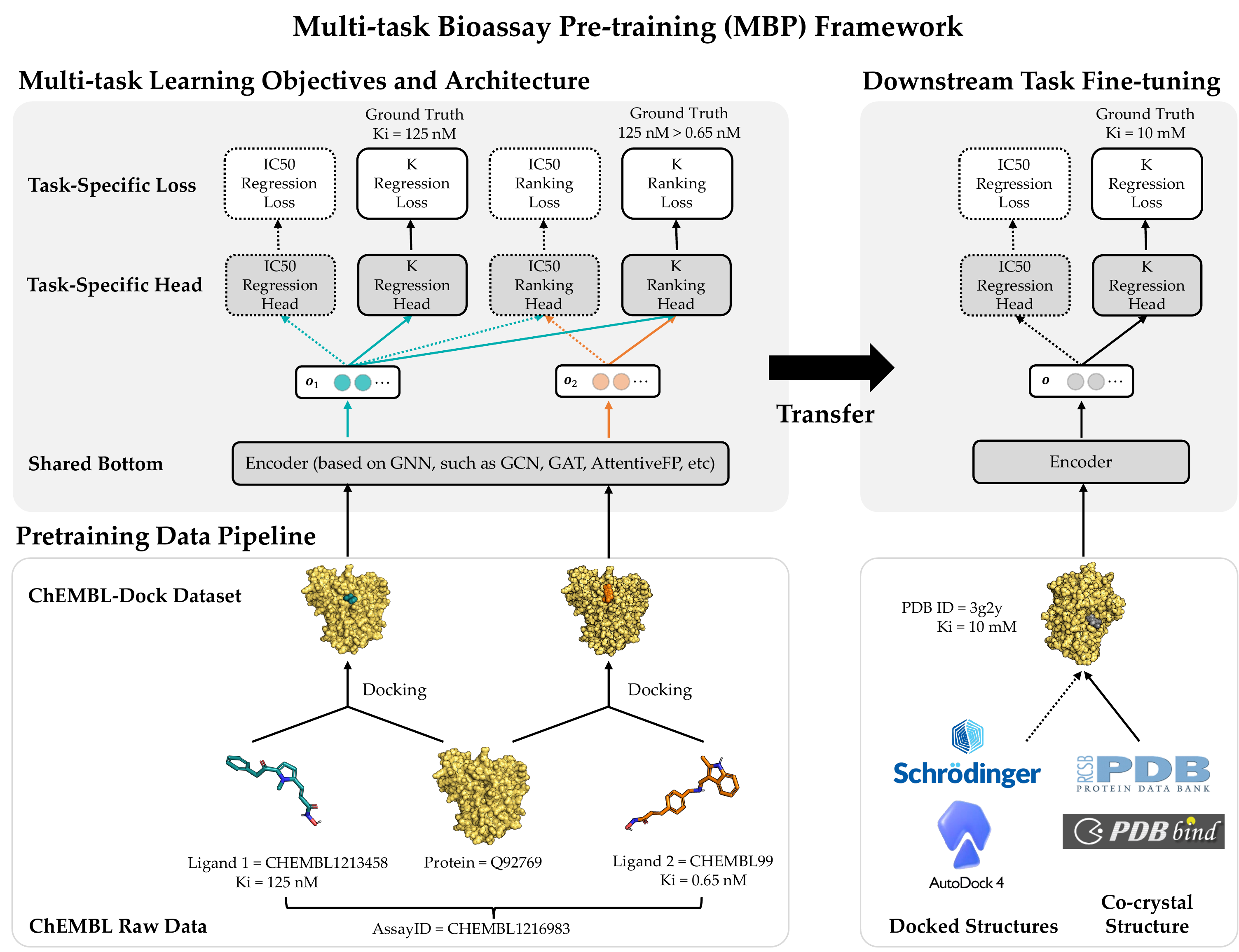}
\vspace{-0.1in}
  \caption{The framework of \modelname in pre-training and fine-tuning. The solid arrows indicate the flow path of the running examples of AssayID = CHEMBL1216983 during pre-training and PDB ID = 3g2y during fine-tuning.}
\vspace{-0.1in}
  \label{fig:3_mt_asrp}
 \end{figure*}

\vspace{-0.05in}
\section{Multi-Task Bioassay Pre-training}
\label{sec:framework}
\vspace{-0.05in}
In this section, we formalize the problem of pre-training for PLBA and then introduce our proposed multi-task bioassay pre-training framework --- \modelname.
After defining the problem, we provide a framework overview of our solution in  \Secref{subsec:framework_overview}. 
Then, in \Secref{subsec:backbone}, we discuss how we develop a GNN-based model as the shared-bottom encoder in multi-task learning.
Finally, we present the curation process of the ChEMBL-Dock dataset used for pre-training in \Secref{subsec:dataset}.

\vpara{Problem Formulation.}
Conceptually, given a protein $P$, a ligand $L$, and the binding conformation $C$ of the ligand to the protein, the problem of structure-based protein-ligand binding affinity prediction is to learn a model $f(P, L, C)$ to predict the binding affinity.
However, due to the rarity and high cost of ground truth 3D structure data, the training of structure-based PLBA prediction models has to be restricted to PDBbind with co-crystal structures. In this work, we aim to leverage the ChEMBL dataset, which contains large-scale protein-ligand binding affinity data but without 3D structures.
As discussed in \Secref{sec:intro}, in order to pre-train a PLBA prediction model on ChEMBL, we have to resolve three challenges, namely label variety, label noise, and missing conformation.


 \vspace{-0.05in}
\subsection{Framework Overview 
}
\label{subsec:framework_overview}
\vspace{-0.05in}
The framework overview of \modelname is illustrated in Fig. \ref{fig:3_mt_asrp}, which includes three main parts --- (1) pre-training data pipeline; (2) multi-task learning objectives and architecture; and (3) downstream task fine-tuning.
In this section, we describe them in detail.

\vpara{Pre-Training Data Pipeline.} 
Before introducing the pre-training data pipeline, we provide the necessary definitions of a bioassay and a  bioassay-specific data pair in \modelname. 

\begin{definition}[A Bioassay in \modelname]
A bioassay is defined as an analytical method to determine the concentration or potency of a substance by its effect on living animals or plants (in vivo) or on living cells or tissues(in vitro) \cite{bliss1957some}. In this work, we mainly focus on bioassays measuring \textit{in vitro} binding of ligands to a protein target. Formally, the $i$-th bioassay is denoted as
\beq{
A_i = \left(P_i, \left\{(L_{ij}, y_{ij})\right\}_{j=1}^{n_i}, t_i\right).
}It means that there are $n_i$ experimental records in bioassay $A_i$, and each record measures the binding affinity $y_{ij}$ of a ligand $L_{ij}$ to the protein target $P_i$. And the type of binding affinity in bioassay $A_i$ is $t_i$, in this work we consider $t_i\in \{\text{Ki}, \text{Kd}, \text{IC50}\}$.
The ChEMBL dataset can be formalized as a collection of bioassays such that $\mathcal{D} = (A_1, A_2, \cdots)$.
\end{definition}

\begin{definition}[A Bioassay-specific Data Pair] 
\label{def:data_sample}
A bioassay-specific data pair is a six-tuple $(P_i, L_{ij}, L_{ik}, y_{ij}, y_{ik}, t_i)$,
indicating there is a bioassay $A_i$
which includes the binding measurement of ligand $L_{ij}$ and ligand $L_{ik}$ to a protein target $P_i$. 
And the experimentally measured  binding affinity (with  type $t_i$) is  $y_{ij}$ and $y_{ik}$, respectively.
\end{definition}

The bioassay-specific data pairs in this work are extracted and randomly sampled from ChEMBL bioassays \cite{Gaulton2011ChEMBLAL}.
We first sample a bioassay $A_i$ with probability proportional to its size, i.e., $\text{Prob}(A_i) \propto n_i$. Then we randomly pick two different ligands $L_{ij}$ and $L_{ik}$ from the sampled assay $A_i$, together with their binding affinity $y_{ij}$ and $y_{ik}$. The above sampling process produces a bioassay-specific data pair $(P_i, L_{ij}, L_{ik}, y_{ij}, y_{ik}, t_i)$ as defined in Definition~\ref{def:data_sample}. Taking the running case shown in \Figref{fig:3_mt_asrp} as an example, the data pair from a ChEMBL bioassay (with AssayId = CHEMBL1216983) can be written as (Q92769, CHEMBL1213458, CHEMBL99, 125nM, 0.65nM, Ki).

Knowledge of the binding conformation of a ligand to a target protein plays a vital role in structure-based drug design, particularly in predicting binding affinity. However, the ground truth co-crystal structure of a protein-ligand complex is experimentally very expensive to determine and is therefore not available in ChEMBL. Consequently, we only have the binding affinities (e.g., $y_{ij}$ and $y_{ik}$) of a ligand to a protein without knowing their conformation and relative orientation. To solve the above missing data problem, we propose to use computationally determined docking poses as an approximation to the true binding conformations. Specifically, we construct a large-scale docking dataset named \datasetname from ChEMBL. For each protein-ligand pair in ChEMBL, we generate its docking poses according to the following three steps. Firstly, we use RDKit library~\cite{greg_landrum_2022_6798971} to generate 3D conformations from the 2D SMILES of the ligand. Then, the 3D structure of a protein is extracted from  PDBbind according to its UniProt ID. Finally, we use docking software SMINA~\cite{Koes2013LessonsLI} to generate the docking poses of the protein-ligand pair. Throughout the rest of this paper, we denote the docking conformation of protein $P_i$ and ligand $L_{ij}$ as $C_{ij}$. 
The detailed data curation process of \datasetname can be found later in \Secref{subsec:dataset}.

\emph{
Overall, 
the pre-training data pipeline generates  bioassay-specific data pairs $(P_i, L_{ij}, L_{ik}, y_{ij}, y_{ik}, t_i)$, retrieves their docking conformations --- $C_{ij}$ and $C_{ik}$ --- from the pre-processed \datasetname datasets, and then feeds them into a multi-task learning model which we will discuss below.
}

\vpara{Multi-task Learning Objectives and Architecture.}
As discussed in \Secref{sec:intro}, there are three main challenges of applying pre-training to the PLBA problem --- missing conformation, label variety, and label noise. 
In this section, we propose to solve the label variety and label noise problem via multi-task learning.

For the label variety challenge, it is intuitive and straightforward
to introduce label-specific tasks for each type of binding affinity measurement.
In this work, we define two categories of label-specific tasks --- IC50 task and K=\{Ki, Kd\} task, which handle bioassay data with affinity measurement type IC50 and Ki/Kd, respectively. Here we merge Ki and Kd as a single task following \cite{Meli2022ScoringFF}, and the main reasons are twofold. Firstly, Ki and Kd are calculated in the same way, except that Kd only considers the physical binding, while Ki specifies the biological effect of this binding to be inhibition. So they can essentially be seen as the same label type. Secondly, the number of Kd data is significantly less compared to Ki data, which may lead to data imbalance if we were to design a separate Kd task.

For the label noise challenge, instead of leveraging learning  with noisy labels techniques~\cite{natarajan2013learning}, we turn to utilize the intrinsic characteristics of bioassay data.
As discussed in \Secref{sec:intro}, the label noise challenge in ChEMBL stems mainly from its data sources and curation process. The binding affinity values from different bioassays were measured under various experimental protocols and conditions (such as temperature and pH value), leading to systematic errors between different assays. However, the binding affinity labels within the same bioassay were usually determined under similar experimental conditions. Thus, intra-bioassay data are more consistent than inter-bioassay ones, and the comparison within a bioassay is much more meaningful. Inspired by the above characteristics of bioassay data, we design both regression tasks and ranking tasks in \modelname{}.
To be more formal, given a bioassay-specific data pair $(P_i, L_{ij}, L_{ik}, y_{ij}, y_{ik}, t_i)$, the regression task
is to directly predict the binding affinity $y_{ij}$, while the ranking task is to compare the binding affinity values within the bioassay, i.e., to classify whether $y_{ij} < y_{ik}$ or  $y_{ij} > y_{ik}$.

In summary, we have $2\times 2=4$ tasks in \modelname{}, namely the IC50 regression task, IC50 ranking task, K regression task, and K ranking tasks. An illustration of these tasks can be found in \Figref{fig:3_mt_asrp}.

As for the multi-task learning architecture, we adopt the shared-bottom technique (also known as the hard parameter sharing) \cite{caruana1993multitask} in \modelname{}. Such a technique 
 shares a bottom encoder among all tasks while keeping several task-specific heads. As illustrated in \Figref{fig:3_mt_asrp}, the model architecture consists of a shared encoder network $f_\text{Enc}$ and four task specific heads --- an IC50 regression head $f_\text{IC50, Reg}$, a K=\{Ki,Kd\} regression head $f_\text{K, Reg}$, an IC50 ranking head $f_\text{IC50, Rank}$ and a K=\{Ki,Kd\} ranking head $f_\text{K, Rank}$. Given a bioassay-specific data pair $(P_i, L_{ij}, L_{ik}, y_{ij}, y_{ik}, t_i)$ together with their conformation $C_{ij}$ and $C_{ik}$ from the pre-training data pipeline, the shared bottom encoder maps them into compact hidden representations shared among tasks:
\beq{
\vo_1 = f_\text{Enc}(P_i, L_{ij}, C_{ij}) \text{ and }
\vo_2 = f_\text{Enc}(P_i, L_{ik}, C_{ik}).
}There are many possibilities for implementing an encoder for protein-ligand complexes, including but not limited to models based on 3D-CNN \cite{Kwon2020AKScoreAP, Jones2020ImprovedPB, Francoeur2020ThreeDimensionalCN}, GNN \cite{Moon2020PIGNetAP, Li2021StructureawareIG, jiang2021interactiongraphnet}, and Transformer \cite{Yan2022GraphsequenceAA, Wang2022ANM, Chen2020TransformerCPIIC}.
In \modelname, we propose a simple and effective shared bottom encoder.
For the sake of clarity, we defer its implementation detail  in \Secref{subsec:backbone}, and focus on multi-task learning in this section. 

For the regression task,
we pick the  task-specific regression head $f_{t_i,\text{Reg}}$ according to the label type $t_i \in \{\text{IC50}, \text{K}\}$ (recall that Ki and Kd have been merged to be a single label type K), and 
predict the binding affinity to be 
$\hat{y}_{ij} = f_{t_i,\text{Reg}}(\vo_1)$. 
The regression loss is calculated using the mean squared error (MSE) loss between ground truth $y_{ij}$ and the predicted value $\hat{y}_{ij}$. More formally, the regression loss is defined as
\beq{
\label{eq:loss_reg}
L_\text{Reg} = \text{MSE} \left(\hat{y}_{ij}, y_{ij}\right).
}It is worth mentioning that for a data pair, only label $y_{ij}$ will be used to compute the regression loss.

Similarly, for the ranking task, 
we select the  task-specific ranking head $f_{t_i,\text{Rank}}$ according to the label type  $t_i \in \{\text{IC50}, \text{K}\}$, 
concatenate the hidden representations as $\vo_1 || \vo_2$, 
and then predict the pairwise ranking to be $\hat{r}_{ijk} = f_{t_i,\text{Rank}}(\vo_1 || \vo_2)$. 
The ranking loss is calculated as the binary cross entropy loss  between ground truth $\mathbb{I}[y_{ij} > y_{ik}]$ and the predicted value $\hat{r}_{ijk}$, where $\mathbb{I}(\cdot)$ denotes the indicator function. More formally, the ranking loss is defined as
\beq{
\label{eq:loss_rank}
L_\text{Rank} = \text{BCE}\left(\hat{r}_{ijk}, \mathbb{I}[y_{ij} > y_{ik}]\right).
}The overall loss function for a
bioassay-specific  data pair is a weighted sum of the regression loss in \Eqref{eq:loss_reg} and ranking loss in \Eqref{eq:loss_rank}:
\beq{
    L_\text{\modelname} = L_\text{Rank} + \lambda  \times L_\text{Reg},
}where $\lambda$ is the weight coefficient for regression loss.

\emph{Overall, we introduce multi-task learning into \modelname{}, aiming to deal with label variety and label noise problems. In the  illustrative example of \modelname{} shown in  \Figref{fig:3_mt_asrp}, 
\modelname{} accepts the bioassay-specific data pair (Q92769, CHEMBL1213458, CHEMBL99, 125nM, 0.65nM, Ki) and their docking poses as inputs, encodes  them to  hidden representations, forwards the K regression head to predict $y_{ij}=125$nM, and also forward the K ranking head to classify $\text{125nM} > \text{0.65nM}$.
}

\vpara{Downstream Task Fine-Tuning.}
The final part of the \modelname framework is the downstream task fine-tuning. 
Given the 3D structure of a protein-ligand complex as input, the downstream task is to predict its binding affinity.  The 3D structure can be either an experimentally determined co-crystal structure or a computationally determined docking pose.
We transfer and fine-tune the shared bottom encoder $f_\text{Enc}$ together with the regression heads 
$f_\text{IC50, Reg}$ and $f_\text{K, Reg}$ in downstream protein-ligand binding affinity datasets (such as PDBbind). The right panel of \Figref{fig:3_mt_asrp} shows how the transferred model predicts the Ki value for a protein-ligand complex from PDBbind (PDB ID=3g2y).

 \vspace{-0.05in}
\subsection{Shared Bottom Encoder}
\label{subsec:backbone}
 \vspace{-0.05in}
For large-scale pre-training, a simple and effective backbone model is of utmost importance. Thus, we design the shared bottom encoder based only on vanilla GNN models.
To simplify, we assume the input of the shared bottom encoder is a 3-tuple $(P, L, C)$, indicating a protein $P$, a ligand $L$, and their binding conformation $C$.

\vpara{Representing Protein-Ligand Complex as Multi-Graphs}
The input protein-ligand complex $(P, L, C)$ is processed into three graphs --- a ligand graph, a protein graph, and a protein-ligand interaction graph. We formally define the three graphs as follows:

\begin{definition}[Ligand Graph]
 A ligand graph, denoted by $\mathcal{G}^L=(\mathcal{V}^L, \mathcal{E}^L)$,  is constructed from the input ligand  $L$. 
 $\mathcal{V}^L$ is the node set where node $i$ represents the $i$-th atom in the ligand.
 Each node $i$ is also associated with (1) atom coordinate $c^L_i$ retrieved from the binding conformation $C$ and (2) atom feature vector $\vx^L_i$. 
 The edge set $\mathcal{E}^L$ is constructed according to the spatial distances among atoms. More formally, the edge set is defined to be \beq{
 \mathcal{E}^L = \left\{(i,j): \norm{c^L_i - c^L_j}{2}< cut^{L}, \forall i, j \in \mathcal{V}^L\right\},
 }where $cut^{L}$ is a distance threshold, and each edge $(i,j)\in \mathcal{E}^L$ is associated with an edge feature vector $\ve^L_{ij}$.
 The node and edge features are obtained by Open Babel \cite{OBoyle2011OpenBA}.
\end{definition}

\begin{definition}[Protein Graph]
 A protein graph, denoted by $\mathcal{G}^P=(\mathcal{V}^P, \mathcal{E}^P)$,  is constructed from the input protein $P$. 
 $\mathcal{V}^P$ is the node set where the node $i$ represents the $i$-th residue in the protein.
 Each node $v^P_i$ is also associated with (1) the alpha carbon coordinate of the $i$-th residue  $c^P_i$ retrieved from the binding conformation $C$ and (2) the residue feature vector $\vx^P_i$. 
 The edge set $\mathcal{E}^P$ is constructed according to the spatial distances among atoms. More formally, the edge set is defined to be \beq{
 \mathcal{E}^P = \left\{(i,j): \norm{c^P_i - c^P_j}{2}< cut^{P}, \forall i, j \in \mathcal{V}^P \right\},
 } where $cut^{P}$ is a distance threshold, and each edge $(i,j)\in \mathcal{E}^P$ is associated with an edge feature vector $\ve^P_{ij}$.
 The  node and edge features are obtained  following \cite{Ganea2021IndependentSM}.
\end{definition}

\begin{definition}[Interaction Graph]
 The protein-ligand interaction graph $\mathcal{G}^{I}=( \mathcal{V}^{P}, \mathcal{V}^{L}, \mathcal{E}^{I})$ is a bipartite graph constructed based on the protein-ligand complex, whose nodes set are the union of protein residues $\mathcal{V}^{P}$ and ligand atoms $\mathcal{V}^{L}$.
 The edge set $\mathcal{E}^{I}$ models the protein-ligand interactions according to spatial distances. More formally,
 \beq{
 \mathcal{E}^{I} = \left\{(i,j):  \norm{c^P_i - c^L_j}{2} < cut^I, \forall i \in \mathcal{V}^P, j \in \mathcal{V}^L
 \right\},
 }where $cut^{I}$ is a spatial distance threshold for interaction, 
 and each edge $(i,j)\in \mathcal{E}^I$ is associated with an edge feature vector $\ve^I_{ij}$.
  The edge features are obtained following \cite{Ganea2021IndependentSM}.
 \end{definition}

When constructing multi-graphs, we follow previous work\cite{Strk2022EquiBindGD, Klicpera2020DirectionalMP, corso2023diffdock, Zhang2022InterResidueDP, Wang2016AccurateDN, Muegge1999AGA, Li2021StructureawareIG} and 
set the distance thresholds as $cut^L=5\mathrm{\mathring{A}}$,  $cut^P=8\mathrm{\mathring{A}}$ and 
 $cut^{I}=12\mathrm{\mathring{A}}$, respectively.
Limited by space, we defer the more detailed multi-graph generation pseudo-code to Appendix~\ref{Construction of Multi-Graph}.

\label{Architecture of Shared Bottom Encoder}
\begin{figure*}
  \centering
  \includegraphics[width=0.9\linewidth]{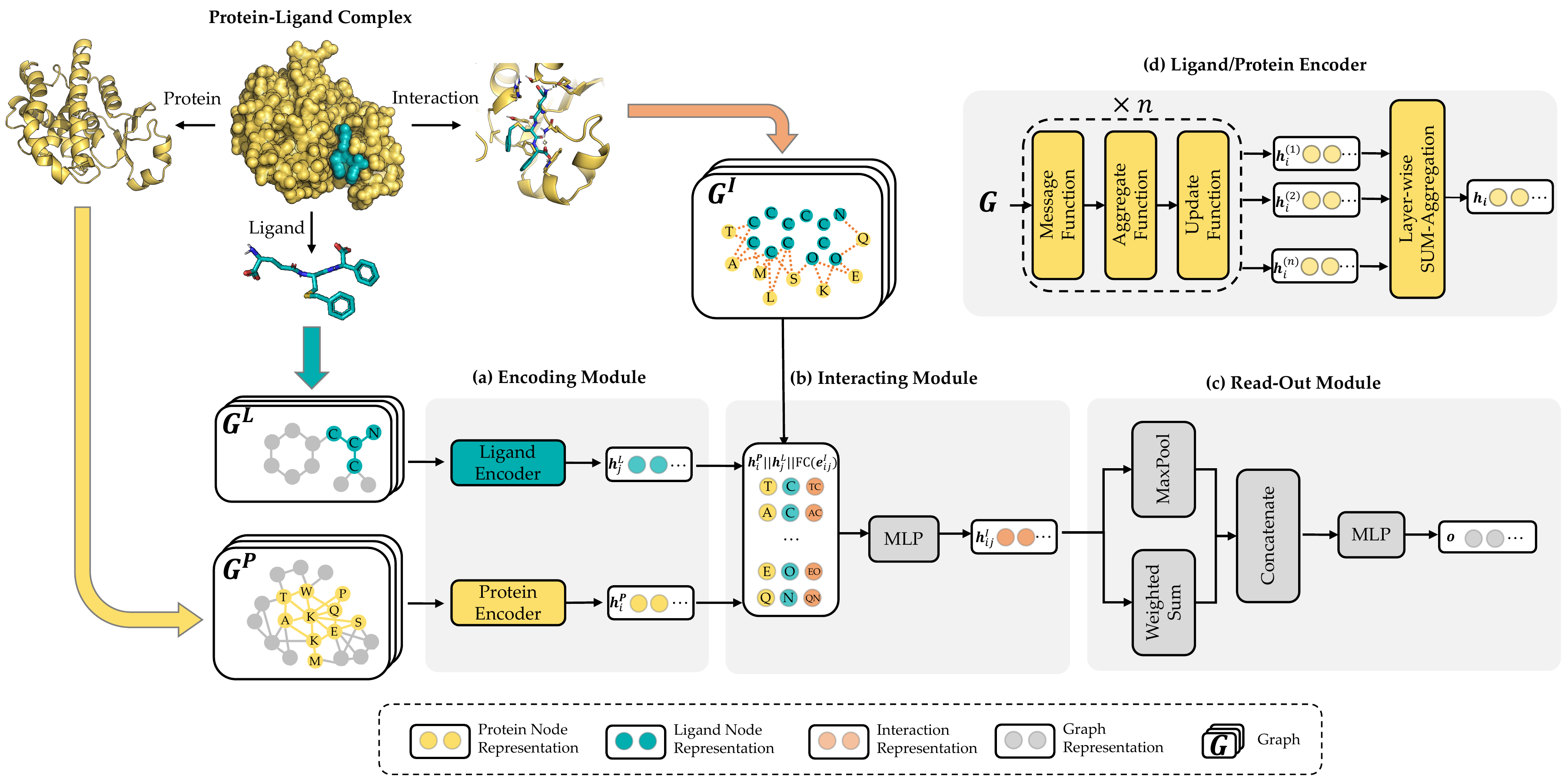}
    \vspace{-0.1in}
  \caption{Shared bottom encoder of \modelname. It contains three modules: (a) encoding module, (b) interacting module, and (c) read-out module. (d) shows the detailed GNN model of the ligand/protein encoder in the encoding module.}
  \label{fig:2_affinity_gnn}
    \vspace{-0.1in}
 \end{figure*}
 
\vpara{Encoding Module~(Ligand/Protein Encoder)}
 Having represented the protein-ligand complex as multi-graphs, we respectively feed the ligand graph $\mathcal{G}^L$ and the protein graph $\mathcal{G}^P$ into the ligand encoder and the protein encoder, aiming to extract informative node representations.
More formally, taking $\mathcal{G}^L$ and $\mathcal{G}^P$  as inputs, we have
\beq{
\mH^L = \text{GNN}(\mathcal{G}^L) \text{ and } \mH^P = \text{GNN}(\mathcal{G}^P).
}Here $\mH^L$ is the ligand embedding matrix of shape $\abs{\mathcal{V}^L} \times d$. And the $i$-th row of $\mH^L$, denoted by $\vh^L_i$, represents the embedding of the $i$-th ligand atom. Similarly, $\mH^P$ is the protein embedding matrix of shape $\abs{\mathcal{V}^P} \times d$. And the $i$-th row of $\mH^P$, denoted by $\vh^P_i$, represents the embedding of the $i$-th protein residue.

 Encoders used here can be any GNN model, such as GCN, GAT, GIN, EGNN, AttentiveFP, etc. 
 Here we briefly review GNNs following the message-passing paradigm following \cite{Gilmer2017NeuralMP} and \cite{Hamilton2017InductiveRL}.
For simplicity and convenience, we assume that the GNN operates on graph $\mathcal{G}$ with node features $\vx_i$ and edge features $\ve_{ij}$, and temporarily ignore whether it is a ligand or protein graph. The message-passing process runs for several iterations. 
At the $\ell$-th iteration, the message-passing is defined according to a message function $M_\ell$, an aggregation function $\text{AGGREGATOR}_\ell$, and an update function $U_\ell$. The embedding $\vh_i^{(\ell)}$ of node $i$ is updated via its message $\vm_i^{(\ell+1)}$: 
\beq{
\label{eq:gnn_message_passing}
\besp{
\vm^{(\ell+1)}_i &= \text{AGGREGATOR}_\ell\left(\left\{M_\ell\left(\vh^{(\ell)}_i,\vh^{(\ell)}_j, \ve_{ij}\right), j \in \mathcal{N}(i)\right\}\right),\\
\vh^{(\ell+1)}_i&= U_\ell \left(\vh^{(\ell)}_i, \vm^{(\ell+1)}_i\right),
}
}where $\mathcal{N}(i)$ is the neighbors of node $i$. Finally, after $n$ iterations of message passing, we sum up the node representations of each layer to get the final node representation, i.e., 
 $\vh_i = \sum_{\ell=1}^{n}{\vh^{(\ell)}_i}$. The GNNs with layer-wise aggregation are also known as jumping knowledge networks~\cite{Xu2018RepresentationLO}.

\vpara{Interacting Module}
After extracting ligand atom embedding $\vh^L_i$ and protein residue embedding $\vh^P_j$ from the  encoding module, the interacting module
is designed to conduct knowledge fusion according to the protein-ligand interaction graph. 
For each protein-ligand interaction edge $(i,j) \in \mathcal{E}^I$, 
 we define its interaction embedding as the concatenation of the protein residue embedding $\vh^P_i$, ligand atom embedding $\vh^L_j$,  and transformed edge features. More formally,
 \beq{
\vh^I_{ij} = \text{MLP}\left(\vh^P_i || \vh^L_j || \text{FC}(\ve^I_{ij})\right),
}where $||$ is the concatenation operator, MLP is a multilayer perceptron, and FC is a fully-connected layer.

\vpara{Read-Out Module}
After obtaining interaction embeddings $\vh^I_{ij}$ for each protein-ligand interaction edge $(i, j) \in \mathcal{E}^{I}$, we further apply an attention-based weighted sum operation to read out a global embedding for the whole protein-ligand complex:
\beq{
\vo_\text{sum} = \sum_{(i,j) \in \mathcal{E}^I} \text{tanh}(\vw^\top \vh^I_{ij})\vh^I_{ij},
}where $\vw$ is the attention vector, $\text{tanh}$ is the hyperbolic tangent function.
Besides, a global maximum pooling operation is adopted to highlight the most informative interaction embedding, s.t., 
$\vo_{\max} = \text{MaxPool}(\vh^I_{ij})$.
We concatenate the above two graph-level embedding to form the final graph embedding for the protein-ligand complex $(P, L, C)$, i.e.,
$\vo = \vo_\text{sum} || \vo_{\max}$.

 \vspace{-0.05in}
\subsection{Pre-training Dataset: \datasetname }
\label{subsec:dataset}
 \vspace{-0.05in}
ChEMBL-Dock is a self-constructed dataset used for pre-training \modelname.
The detailed ChEMBL-based curation workflow, including the data collection \& cleaning step and molecular docking step, can be found in Appendix~\ref{Curation workflow of dataset}.
Here, we compare it to other commonly used protein-ligand complex datasets in terms of the label, 3D structure, protein diversity, molecular diversity, and dataset size in Fig.S1 and Table~\ref{tbl:dataset_comparison_2}.

\begin{wraptable}{r}{0.5\textwidth}
\caption{Overview of data represented in PDBbind, CrossDocked, and \datasetname, respectively.}
\scalebox{0.75}{
\label{tbl:dataset_comparison_2}
\small
\begin{tabular}{l|c|c|c}
\toprule
& \textbf{PDBbind} & \textbf{CrossDocked} & \textbf{\datasetname}\\
\midrule
Protein
& 3,890 & 2,922 & 963 \\
Ligand
& 15,193 & 13,780 & 200,728 \\
Protein-ligand pair
& 19,443 & / & 313,224\\
Pose
& 19,443 &  22,584,102 & 2,819,016 \\
Assay
& / &  / & 21,686 \\
\bottomrule
\end{tabular}
}
\end{wraptable}

Combining the strengths of molecular docking and ChEMBL, \datasetname provides a large-scale 3D protein-ligand complex dataset with corresponding experimental affinity labels.
While the quality of the docked 3D structures of the complexes in \datasetname is not as high as that of the 3D co-crystal structures of the protein-ligand complexes in PDBbind, \datasetname provides a much larger number of 3D structures of protein-ligand complexes than the PDBbind database.
By comparing \datasetname and CrossDocked, two datasets generated through molecular docking, it is evident that \datasetname exhibits a higher molecular diversity than CrossDocked, suggesting its potential to provide a more comprehensive dataset for drug discovery research.

\vspace{-0.1in}
\section{Experiments}
\label{sec:exp}
\vspace{-0.1in}
\subsection{Experimental Setup}
\vspace{-0.05in}
\vpara{Downstream Datasets.}
 Two publicly available datasets are used to comprehensively evaluate the performance of models.
 \begin{itemize}[leftmargin=*]
 \vspace{-0.05in}
\item \textbf{PDBbind v2016}~\cite{Liu2017ForgingTB} is a famous benchmark for evaluating the performance of models in predicting PLBA.
 The dataset includes three overlapping subsets: the general set (13,283 3D protein-ligand complexes), the refined set (4,057 complexes selected out of the general set with better quality), and the core set (290 complexes selected as the highest quality benchmark for testing). We refer to the difference between the refined and core subsets (3,767 complexes) as the refined set for convenience.
 The general set contains IC50 data and K data, while the refined set and core only contain K data. In this paper, the core set is used as the test set, and we train models on the refined set or the general set. 
 \vspace{-0.04in}
 \item \textbf{CSAR-HiQ}~\cite{Dunbar2013CSARDS} is a publicly available dataset of 3D protein-ligand complexes with associated experimental affinity labels. 
 Data included in CSAR-HiQ are K data.
 When training models on the refined set of PDBbind, CSAR-HiQ is typically used to evaluate the generalization performance of the model\cite{Li2021StructureawareIG}.
 In this paper, we create an independent test set of 135 samples based on CSAR-HiQ by removing samples that already exist in the PDBbind v2016 refined set.
\end{itemize}

 \vpara{Baselines.}
We mainly compare \modelname with four families of methods.
The first family is machine learning-based methods such as Linear Regression (LR), Support Vector Regression (SVR), and RF-Score \cite{Ballester2010AML}.
The second family is CNN-based methods, including Pafnucy \cite{StepniewskaDziubinska2018DevelopmentAE} and OnionNet \cite{Zheng2019OnionNetAM}.
The third family of baselines is GraphDTA \cite{nguyen2021graphdta} methods, including GCN, GAT, GIN, GAT-GCN.
The fourth family of baselines is GNN-based methods containing SGCN \cite{danel2020spatial}, GNN-DTI \cite{lim2019predicting}, DMPNN \cite{yang2019analyzing}, MAT \cite{Maziarka2020MoleculeAT}, DimeNet \cite{Gasteiger2020Directional}, CMPNN \cite{Song2020CommunicativeRL}, IGN \cite{jiang2021interactiongraphnet}, and SIGN \cite{Li2021StructureawareIG}.
In addition, the recent molecular docking method TANKBind\cite{lu2022tankbind} and molecular pre-training method Transformer-M\cite{Luo2022OneTC} are also compared as they have been applied to perform PLBA predictions.

 \vpara{Evaluation Metrics.}
 Root Mean Square Error (RMSE), Mean Absolute Error (MAE), Standard Deviation (SD), and Pearson’s correlation coefficient (R) are used to evaluate the performance of PLBA prediction \cite{Li2021StructureawareIG}.
 The definition of these metrics can be found in Appendix~\ref{Experiment Details}.
 
 \vpara{Training Parameter Settings.} 
The models were trained using Adam \cite{Kingma2015AdamAM} with an initial learning rate of $10^{-3}$ and an $L_{2}$ regularization factor of $10^{-6}$.
The learning rate was scaled down by 0.6 if no drop in training loss was observed for 10 consecutive epochs.
For pre-training, the number of training epochs was set to 100, while for fine-tuning, the number of training epochs was set to 1000 with an early stopping rule of 70 epochs if no improvement in the validation performance was observed.

 \vspace{-0.1in}
\subsection{Experimental Results}
\label{sec:overall_exmperiments}
\vspace{-0.05in}

In this work, we employ five different GNNs in the shared bottom encoder of \modelname, which are denoted as \modelname-X (where X corresponds to the GNN used) for distinction. 
For example, \modelname-GCN denotes the \modelname model using GCN in its shared bottom encoder. 
Unless specified otherwise, AttentiveFP is used as the default GNN in \modelname.

\begin{figure*}
  \centering
  \includegraphics[width=\linewidth]{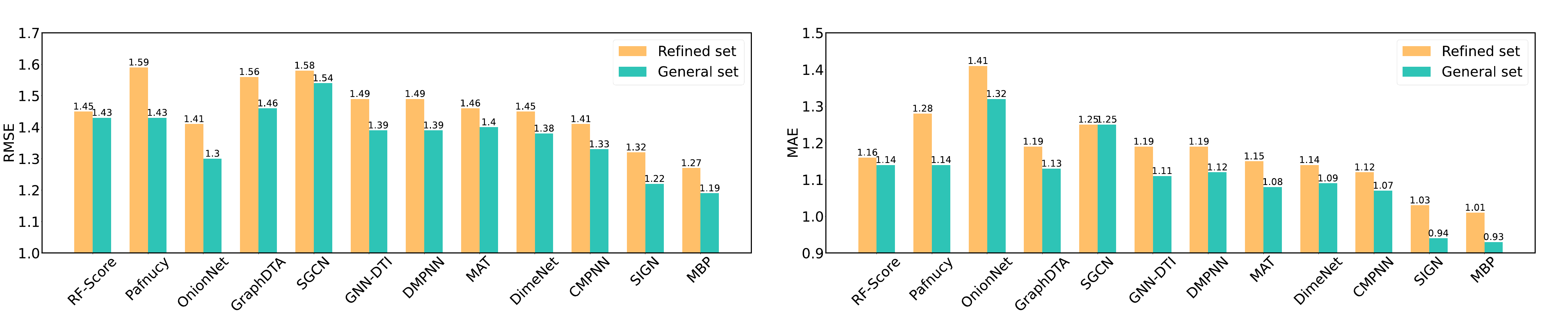}
    \vspace{-0.3in}
  \caption{Performance improvements of baselines and \modelname on the PDBbind benchmark when training on general set.}
  \label{fig:8_general_set}
 \end{figure*}

\vpara{Overall Performance Comparison on PDBbind core set and CSAR-HiQ.}
We first fine-tune \modelname on the PDBbind refined set and report the test performance averaged over five repetitions for each method on the PDBbind core set and CSAR-HiQ set in Table \ref{tbl:sota_comparison}.
It can be observed that \modelname achieves the best performance across all metrics of the two publicly available datasets.
In particular, \modelname-AttentiveFP and \modelname-EGNN outperforms all competing methods on the PDBbind core set. Compared with SIGN, which the previous SOTA method \cite{Li2021StructureawareIG}, \modelname-AttentiveFP achieving an improvement of 4.0\%, 2.7\%, 6.3\%, and 3.5\% on RMSE, MAE, SD, and R, respectively. 
And on the CSAR-HiQ dataset, \modelname-AttentiveFP also achieves results superior to the other competing methods. For instance, it attains more than 6.3\%, 6.6\%, 10.1\%, and 4.9\% on RMSE, MAE, SD, and R gain compared to SIGN.
Both \modelname-EGNN and \modelname-AttentiveFP surpass the previous best method, SIGN, and are the best two methods in the current results. In addition, it is worth noting that \modelname with even the simplest GNN models (e.g., GCN) outperforms SIGN on the CSAR-HiQ dataset.
This indicates that the proposed multi-task pre-training framework is able to improve the capacity and generalization of the backbone model in the PLBA prediction problem. To further evaluate the generalization performance of the proposed model, we conduct an extra experiment on the PDBbind general set. As shown in \Figref{fig:8_general_set}, comparing to all baselines, \modelname achieves the best performance in terms of both RMSE and MAE.

For the molecular docking method TANKBind and the molecular pre-training method Transformer-M, we conducted experiments by adhering to the reported settings of these methods and fine-tuned our MBP model accordingly to ensure a fair comparison. Due to the variations in experimental settings and limited space, we provide a detailed comparison setting and results in Appendix~\ref{Additional experimental results}. The outcomes presented in Table S1 and Table S2 demonstrate the effectiveness and competitiveness of MBP, even when compared to these powerful methods.

\begin{table*}
  \centering

\caption{Test performance comparison on the PDBbind v2016 core set and the CSAR-HiQ dataset. The mean RMSE, MAE, SD, and R (std) over 3 repetitions are reported. The best two results are highlighted in \textbf{bold}.}
\label{tbl:sota_comparison}
\scalebox{0.6}{
\begin{threeparttable}
\begin{tabular}{@{}ll|cccc|cccc@{}}
\toprule
\multicolumn{2}{c|}{\multirow{2}{*}{Method}}&
\multicolumn{4}{c|}{PDBbind core set} & \multicolumn{4}{c}{CSAR-HiQ dataset}\\
\cline{3-10}
& & RMSE$\downarrow$ & MAE$\downarrow$ & SD$\downarrow$ & R$\uparrow$ & RMSE$\downarrow$ & MAE$\downarrow$ & SD$\downarrow$ & R$\uparrow$ \\
\midrule
\multirow{3}{1.5cm}{ML-based} 
& LR 
& 1.675 (0.000) & 1.358 (0.000) & 1.612 (0.000) & 0.671 (0.000)
& 2.071 (0.000) & 1.622 (0.000) & 1.973 (0.000) & 0.652 (0.000)\\
& SVR 
& 1.555 (0.000) & 1.264 (0.000) & 1.493 (0.000) & 0.727 (0.000)
& 1.995 (0.000) & 1.553 (0.000) & 1.911 (0.000) & 0.679 (0.000)\\
& RF-Score 
& 1.446 (0.008) & 1.161 (0.007) & 1.335 (0.010) & 0.789 (0.003) 
& 1.947 (0.012) & 1.466 (0.009) & 1.796 (0.020) & 0.723 (0.007)\\
\midrule
\multirow{2}{1.5cm}{CNN-based} 
& Pafnucy 
& 1.585 (0.013) & 1.284 (0.021) & 1.563 (0.022) & 0.695 (0.011) 
& 1.939 (0.103) & 1.562 (0.094) & 1.885 (0.071) & 0.686 (0.027)\\
& OnionNet 
& 1.407 (0.034) & 1.078 (0.028) & 1.391 (0.038) & 0.768 (0.014)
& 1.927 (0.071) & 1.471 (0.031) & 1.877 (0.097) & 0.690 (0.040)\\
\midrule
\multirow{4}{1.5cm}{GraphDTA}
& GCN
& 1.735 (0.034) & 1.343 (0.037) & 1.719 (0.027) & 0.613 (0.016) 
& 2.324 (0.079) & 1.732 (0.065) & 2.302 (0.061) & 0.464 (0.047)\\ 
& GAT
& 1.765 (0.026) & 1.354 (0.033) & 1.740 (0.027) & 0.601 (0.016)
& 2.213 (0.053) & 1.651 (0.061) & 2.215 (0.050) & 0.524 (0.032)\\ 
& GIN 
& 1.640 (0.044) & 1.261 (0.044) & 1.621 (0.036) & 0.667 (0.018)
& 2.158 (0.074) & 1.624 (0.058) & 2.156 (0.088) & 0.558 (0.047)\\ 
& GAT-GCN 
& 1.562 (0.022) & 1.191 (0.016) & 1.558 (0.018) & 0.697 (0.008)
& 1.980 (0.055) & 1.493 (0.046) & 1.969 (0.057) & 0.653 (0.026)\\ 
\midrule
\multirow{8}{1.5cm}{GNN-based}
& SGCN 
& 1.583 (0.033) & 1.250 (0.036) & 1.582 (0.320) & 0.686 (0.015)
& 1.902 (0.063) & 1.472 (0.067) & 1.891 (0.077) & 0.686 (0.030)\\ 
& GNN-DTI 
& 1.492 (0.025) & 1.192 (0.032) & 1.471 (0.051) & 0.736 (0.021) 
& 1.972 (0.061) & 1.547 (0.058) & 1.834 (0.090) & 0.709 (0.035)\\ 
& DMPNN 
& 1.493 (0.016) & 1.188 (0.009) & 1.489 (0.014) & 0.729 (0.006) 
& 1.886 (0.026) & 1.488 (0.054) & 1.865 (0.035) & 0.697 (0.013)\\ 
& MAT 
& 1.457 (0.037) & 1.154 (0.037) & 1.445 (0.033) & 0.747 (0.013) 
& 1.879 (0.065) & 1.435 (0.058) & 1.816 (0.083) & 0.715 (0.030)\\ 
& DimeNet 
& 1.453 (0.027) & 1.138 (0.026) & 1.434 (0.023) & 0.752 (0.010) 
& 1.805 (0.036) & 1.338 (0.026) & 1.798 (0.027) & 0.723 (0.010)\\ 
& CMPNN 
& 1.408 (0.028) & 1.117 (0.031) & 1.399 (0.025) & 0.765 (0.009)
& 1.839 (0.096) & 1.411 (0.064) & 1.767 (0.103) & 0.730 (0.052)\\ 
& IGN
& 1.519 (0.055) & 1.187 (0.042) & 1.513 (0.052) & 0.718 (0.023) 
& 2.051 (0.077) & 1.604 (0.043) & 1.834 (0.095) & 0.607 (0.057)\\ 
& SIGN 
& 1.316 (0.031) & 1.027 (0.025) & 1.312 (0.035) & 0.797 (0.012)
& 1.735 (0.031) & 1.327 (0.040) & 1.709 (0.044) & 0.754 (0.014) \\
\midrule
\multirow{5}{1.5cm}{\textbf{Ours}}
& \modelname-GCN
& 1.333 (0.019) & 1.056 (0.011) & 1.306 (0.018) & 0.800 (0.006)
& 1.718 (0.044) & 1.348 (0.025) & 1.659 (0.048) & 0.751 (0.017)\\
& \modelname-GIN
& 1.375 (0.031) & 1.088 (0.026) & 1.352 (0.044) & 0.783 (0.016)
& 1.748 (0.028) & 1.391 (0.026) & 1.664 (0.050) & 0.749 (0.017) \\
& \modelname-GAT
& 1.393 (0.017) & 1.122 (0.007) & 1.367 (0.023) & 0.778 (0.008) 
& 1.703 (0.039)& 1.317 (0.041) & 1.606 (0.022) & 0.769 (0.007)\\
& \modelname-EGNN
& \textbf{1.298 (0.029)} & \textbf{1.023 (0.025)} & \textbf{1.262 (0.033)} & \textbf{0.815 (0.011)}
& \textbf{1.649 (0.045)} & \textbf{1.242 (0.036)} & \textbf{1.548 (0.055)} & \textbf{0.788 (0.016)}\\
& \modelname-AttentiveFP
& \textbf{1.263 (0.023)} & \textbf{0.999 (0.024)} & \textbf{1.229 (0.026)} & \textbf{0.825 (0.008)} 
& \textbf{1.624 (0.037)} & \textbf{1.240 (0.038)} & \textbf{1.536 (0.052)} & \textbf{0.791 (0.016)} \\ 
\bottomrule
\end{tabular}
\begin{tablenotes}
 \item[1] The result of IGN was obtained by repeating the protocol provided by its authors. The other results were taken from \cite{Li2021StructureawareIG}.
\end{tablenotes}
\end{threeparttable}
}
\vspace{-0.2in}

\end{table*}

\vspace{-0.1in}
\subsection{Ablation Studies}
\vspace{-0.05in}

In this section, we conduct extensive ablation studies to investigate the role of different components in \modelname. In ablation studies, all \modelname models are fine-tuned on the PDBbind refined set.

\vpara{Multi-Task Learning Objectives.}
We perform an ablation study to investigate the effect of multi-task learning. Table~\ref{tbl:ablation_study} shows the results of our \modelname with different learning tasks. We have two main observations:
\begin{itemize}[leftmargin=*]
    \vspace{-0.08in}
    \item \textbf{Regarding regression tasks and ranking tasks}, we find that on both PDBbind core set and independent CSAR-HiQ set, \modelname with a combination of both regression and ranking tasks can always outperform \modelname with only regression or ranking tasks, indicating the power of ensembling regression and bioassay-specific ranking.
    \vspace{-0.03in}
    \item \textbf{Regarding IC50 tasks and K tasks}, we also find that \modelname pre-trained with both IC50 and K tasks is better than that using only IC50 or K tasks. This implies that \modelname is able to learn the task correlation between IC50 and K data from ChEMBL and transfer the knowledge to the PDBbind core set, which only contains Ki/Kd data.
\end{itemize}
\vspace{-0.08in}
These results justify the effectiveness of the multi-task learning objectives designed in \modelname.

\begin{table*}
\caption{Ablation study of \modelname with different pre-training tasks. 
The mean RMSE, MAE, SD, and R (std) over 5 repetitions are reported. The best two results are highlighted in \textbf{bold}.}
\label{tbl:ablation_study}
\scalebox{0.62 }{
\begin{tabular}{@{}cc|cc|cccc|cccc@{}}
\toprule
\multicolumn{2}{c|}{Regression} & \multicolumn{2}{c|}{Ranking} & 
\multicolumn{4}{c|}{PDBbind core set} & \multicolumn{4}{c}{CSAR-HiQ set}\\
\cmidrule{5-12}
IC50 & K  & IC50 & K  & RMSE$\downarrow$ & MAE$\downarrow$ & SD$\downarrow$ & R$\uparrow$ & RMSE$\downarrow$ & MAE$\downarrow$ & SD$\downarrow$ & R$\uparrow$ \\
\midrule
&&&
& 1.377 (0.045)	& 1.075 (0.040)	& 1.366 (0.042) & 0.778 (0.015)
& 1.661 (0.028) & 1.270 (0.035) & 1.629 (0.043) & 0.777 (0.008)\\
\midrule
\checkmark&&&
& 1.364 (0.009) & 1.077 (0.005) & 1.351 (0.010) & 0.784 (0.004)
& 1.693 (0.038) & 1.293 (0.031) & 1.628 (0.036) & 0.761 (0.011) \\
&&\checkmark&
& 1.418 (0.036) & 1.120 (0.041) & 1.398 (0.037) & 0.766 (0.014)
& 1.787 (0.082) & 1.386 (0.062) & 1.756 (0.062) & 0.714 (0.024) \\
\checkmark&&\checkmark&
& 1.315 (0.011)	& 1.055 (0.010)	& 1.268 (0.014) & 0.813 (0.005)
& 1.690 (0.037) & 1.268 (0.048) & \textbf{1.470 (0.052)} & 0.764 (0.016)\\
\midrule
&\checkmark&&
& 1.292 (0.025)	& 1.018 (0.023)	& 1.267 (0.032) & 0.813 (0.011)
& 1.704 (0.132) & 1.254 (0.053) & 1.586 (0.060) & 0.746 (0.058)\\
&&&\checkmark
& 1.372 (0.029) & 1.096 (0.032) & 1.365 (0.036) & 0.779 (0.013)
& 1.659 (0.008) & 1.294 (0.026) & 1.601 (0.032) & 0.771 (0.010)\\
&\checkmark&&\checkmark
& \textbf{1.283 (0.023)} & \textbf{1.017 (0.014)} & 1.255 (0.032) & \textbf{0.817 (0.010)}
& \textbf{1.637 (0.018)} & \textbf{1.233 (0.036)} & 1.580 (0.008) & \textbf{0.788 (0.011)} \\
\midrule
\checkmark&\checkmark&&
& 1.287 (0.025) & 1.027 (0.024) & \textbf{1.254 (0.031)} & \textbf{0.817 (0.010)}
& 1.662 (0.048) & 1.269 (0.032) & 1.544 (0.058) & 0.789 (0.018) \\
&&\checkmark&\checkmark
& 1.325 (0.010) & 1.048 (0.011) & 1.307 (0.013) & 0.800 (0.044)
& 1.674 (0.045) & 1.293 (0.033) & 1.616 (0.049) & 0.766 (0.016) \\
\checkmark&\checkmark&\checkmark&\checkmark
& \textbf{1.263 (0.023)} & \textbf{0.999 (0.024)} & \textbf{1.229 (0.026)} & \textbf{0.825 (0.008)} 
& \textbf{1.624 (0.037)} & \textbf{1.240 (0.038)} & \textbf{1.536 (0.052)} & \textbf{0.791 (0.016)} \\ 
\bottomrule
\end{tabular}
}
\vspace{-0.25in}

\end{table*}

\vpara{GNN Used in Shared Bottom Encoder.}
As shown in Table \ref{tbl:sota_comparison}, we benchmark and compare \modelname with different GNNs in the shared bottom encoder. 
We choose five popular GNN models --- GCN, GIN, GAT, EGNN, and AttentiveFP.
EGNN and AttentiveFP are able to capture the 3D structure of biomolecules, while GCN, GIN, and GAT are mainly designed for general graphs which can not capture structural information directly.
We have two interesting observations. Firstly, all GNN models, even the vanilla GCN, achieve comparable or better performance than previous methods. For example, MBP-GCN achieves RMSE of 1.718 on the CSAR-HiQ set, slightly better than SIGN's 1.735. Secondly, GNNs that explicitly capture 3D structure information (EGNN and AttentiveFP) outperform GNNs designed for general graphs (GCN, GAT, GIN).

 \begin{wrapfigure}{r}{0.5\linewidth}
  \centering
  \includegraphics[width=\linewidth]{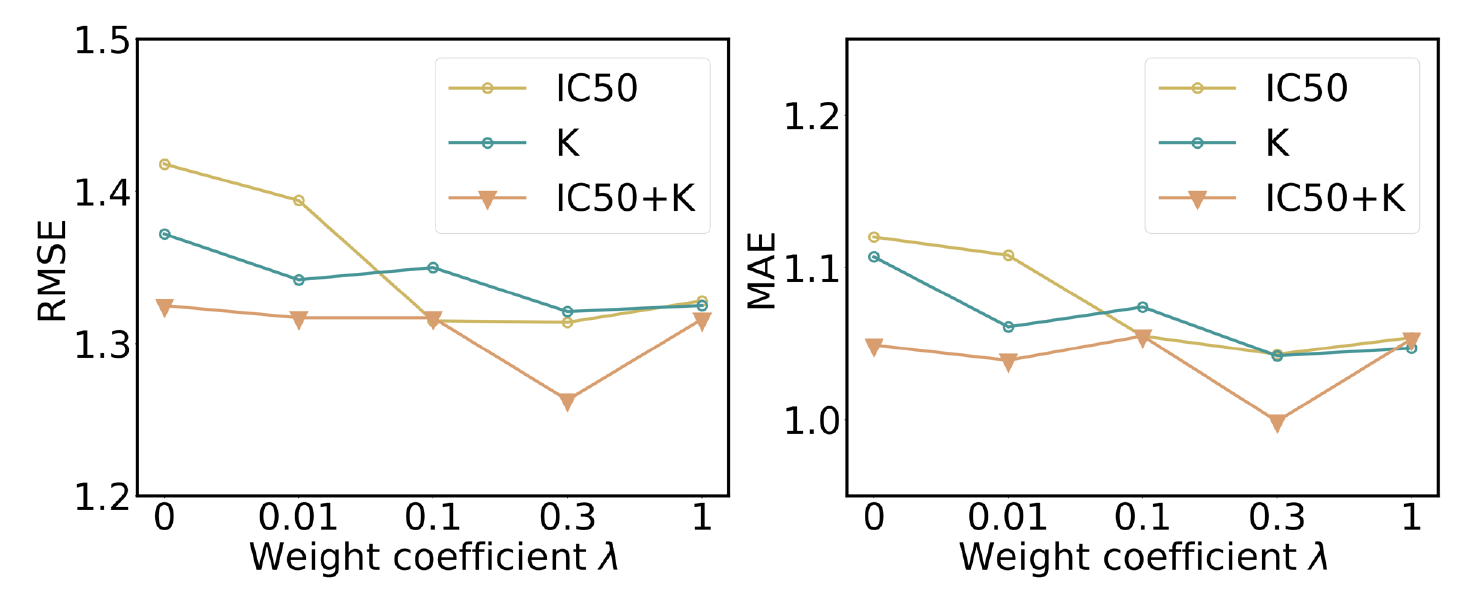}
    \vspace{-0.2in}
  \caption{
  Test RMSE and MAE of \modelname on the PDBbind core set with varying weight coefficients $\lambda$ of the regression loss.}
  \label{fig:7_hyper_tune}
   \vspace{-0.1in}
 \end{wrapfigure}

\vpara{Hyper-Parameter Studies.}
We ablate the weight coefficient $\lambda$ of regression loss in \modelname, which is crucial to the performance of  \modelname.
Intuitively, too small a  $\lambda$ may hurt the ability to predict binding affinity, while too large a $\lambda$ may aggravate the label noise problem.
We vary the weight coefficients $\lambda$ from $\{0, 0.01, 0.1, 0.3, 1.0\}$, and then depict the tendency curves of the test RMSE w.r.t. $\lambda$ in \Figref{fig:7_hyper_tune}. As expected, too large or too small a weight coefficient leads to worse performance in different multi-task settings (i.e., IC50 tasks, K tasks, and IC50+K tasks).

\vspace{-0.1in}
\section{Conclusion}
\label{sec:conclusion}
\vspace{-0.05in}
Protein-ligand binding affinity (PLBA) is a critical measure in drug discovery. However, the limited availability of high-quality training data and the variety and noise of PLBA labels pose difficulties in improving the performance of these models through pre-training. In this paper, we introduce the \datasetname dataset, which contains 313,224 3D protein-ligand complexes with experimental PLBA labels. Based on this dataset, we proposed the \modelname method that addresses the label variety and noise problems through multi-task learning and assay-specific ranking tasks.

\newpage
\bibliographystyle{unsrt}
\bibliography{sample-base}

\appendix
\clearpage
\setcounter{figure}{0}
\setcounter{table}{0}
\makeatletter 
\renewcommand{\thefigure}{S\@arabic\c@figure}
\renewcommand{\thetable}{S\@arabic\c@table}
\makeatother
\renewcommand{\appendixname}{Appendix~\Alph{section}}

\section{Pre-training and fine-tuning procedures}
 In this section, to facilitate other researchers to reproduce our results, we provide the pseudocode of \modelname's pre-training procedure and fine-tuneing procedure in Algorithm \ref{alg1} and Algorithm \ref{alg2}.

 \begin{algorithm} 
\small
	\caption{Pre-training Procedure for \modelname.} 
	\label{alg1} 
    \KwIn{Pre-training set $\mathcal{D}$}
    \KwOut{Pre-trained model parameters $\theta$}
    Randomly initialize the parameter $\theta$ of \modelname\;
    \For {iteration = 1,2,..,100}
    {
        \For{each batch from training samples do}
        {
            Obtain protein node representations $\vh^P$ using protein encoder\;
            Obtain ligand node representations $\vh^L$ using ligand encoder\;
            Obtain interaction embedding $\vh^I$ \;
            Obtain graph-level embedding $\vo$ \;
            Calculate the affinity prediction $\hat{y}$ using $\vo$\;
            Calculate the ranking prediction $\hat{r}$ using $\vo$\;
            Calculate \modelname overall $L_{\modelname}$\;
            Update parameters $\theta$ according to the gradient of $L_{\modelname}$\;
        }
    }
    \Return $\theta$
\end{algorithm}

\begin{algorithm} 
\small
	\caption{Fine-tuning Procedure for \modelname.} 
	\label{alg2} 
    \KwIn{Finetuning set $\mathcal{D}$, Pre-trained model parameters $\theta$}
    \KwOut{Finetuned model parameters $\theta$}
    \For {iteration = 1,2,..,1000}
    {
        \For{each batch from training samples do}
        {
            Obtain protein node representations $\vh^P$ using protein encoder\;
            Obtain ligand node representations $\vh^L$ using ligand encoder\;
            Obtain interaction embedding $\vh^I$ \;
            Obtain graph-level embedding $\vo$ \;
            Calculate the affinity prediction $\hat{y}$ using $\vo$\;
            Calculate the regression loss $L_{Reg}$\;
            Update parameters $\theta$ according to the gradient of $L_{Reg}$\;
        }
    }
    \Return $\theta$
\end{algorithm}

\section{Construction of Multi-Graph}
\label{Construction of Multi-Graph}
Protein graphs and ligand graphs only contain intramolecular connections, and intermolecular connections are available in protein-ligand interaction graphs.
We present the pseudocode of the construction process of multi-graph in Algorithm~\ref{alg3}.

\begin{algorithm} 
\small
	\caption{Construction of Multi-Graph} 
	\label{alg3} 
    \KwIn{Protein-ligand complex $(P,L,C)$, intramolecular cutoff distance $cut^L$, $cut^P$ and intermolecular cutoff distance $cut^I$}
    \KwOut{The Multi-Graph $\mathcal{G}_m=(\mathcal{G}^L,\mathcal{G}^P,\mathcal{G}^I)$}
    
    Initialize $\mathcal{E}^L \leftarrow \left\{\right\}$\ and extract $\mathcal{V}^L $ from ligand $L$;
    
    \For {node pair ($v_i, v_j \in \mathcal{V}^L \times \mathcal{V}^L$)}
    {
        Calculate distance $d_{ij} \leftarrow \norm{c_i- c_j}{2}$\;
        \If{$d_{ij} \leq cut^L$}
            {
                Update edge set $\mathcal{E}^L \leftarrow \mathcal{E}^L \cup (i,j) $\;
            }
    }
    
    Initialize $\mathcal{E}^P \leftarrow \left\{\right\}$\ and extract $\mathcal{V}^P $ from protein $P$;
    
    \For {node pair ($v_i, v_j \in \mathcal{V}^P \times \mathcal{V}^P$)}
    {
        Calculate distance $d_{ij} \leftarrow \norm{c_i - c_j}{2}$\;
        \If{$d_{ij} \leq cut^P$}
        {
            Update edge set $\mathcal{E}^P \leftarrow \mathcal{E}^P \cup (i,j)$\;
        }
    }
    
    Initialize $\mathcal{E}^I \leftarrow \left\{\right\}$\ and extract $\mathcal{V}^I $ from ligand $L$ and protein $P$;
    
    \For {node pair ($v_i, v_j \in \mathcal{V}^L \times \mathcal{V}^P$)}
    {
        Calculate distance $d_{ij} \leftarrow \norm{c_i-c_j}{2}$\;
        \If{$d_{ij} \leq cut^I$ }
        {
            Update edge set $\mathcal{E}^I \leftarrow \mathcal{E}^I \cup (i,j) $\;
        }
    }
    $\mathcal{G}^L \leftarrow (\mathcal{V}^L,\mathcal{E}^L)$\;
    $\mathcal{G}^P \leftarrow (\mathcal{V}^P,\mathcal{E}^P)$\;
    $\mathcal{G}^I \leftarrow (\mathcal{V}^I,\mathcal{E}^I)$\;
    \Return $\mathcal{G}^L,\mathcal{G}^P,\mathcal{G}^I$
\end{algorithm}

\section{Experiment Details}
\label{Experiment Details}
\subsection{Hardwares}
We pre-trained our model on four NVIDIA Tesla V100 GPUs (32G) and fine-tune it on a single GPU.

\subsection{Parameter settings}
When constructing multi-graph input, we set the intramolecular cutoff distance for ligand $cut^L$ to 5.0$\mathrm{\mathring{A}}$ and intramolecular cutoff distance for protein  $cut^P$ to 8.0 $\mathrm{\mathring{A}}$, with an intermolecular cutoff distance $cut^I$=12.0$\mathrm{\mathring{A}}$.
For GNN used in the shared bottom encoder of \modelname, we use five different GNNs: GCN, GIN, GAT, EGNN, and AttentiveFP.
The numbers of message-passing layers in GCN, GIN, GAT, EGNN, and AttentiveFP are set to 2, 3, 3, 3, 3, respectively.
Lengths of node representations learned by the ligand encoder and protein encoder are both set to 128.
For the auxiliary task, we set the balancing coefficient to 0.1.
We use a three-layer MLP to predict PLBA and a one-layer MLP for auxiliary task prediction. The hidden layer dimensions of these two MLPs are 128.

\subsection{Evaluation Metrics}
Here, we give the formal formulas of the evaluation metrics mentioned in the experiments section:
\begin{equation}
    RMSE=\sqrt{\frac{1}{|D|}\sum_{i=1}^{|D|}(\hat{y_i}-y_i)^2},
\end{equation}
\begin{equation}
    MAE=\frac{1}{|D|}\sum_{i=1}^{|D|}|\hat{y_i}-y_i|,
\end{equation}
\begin{equation}
    SD=\sqrt{\frac{1}{|D|-1}\sum_{i=1}^{|D|}[y_i-(a+b\hat{y_i})]^2},
\end{equation}
\begin{equation}
    R=\frac{ \sum_{i=1}^{|D|} (\hat{y_i}-\bar{\hat{y_i}}) (y_i-\bar{y_i}) }{\sqrt{\sum_{i=1}^{|D|} (\hat{y_i}-\bar{\hat{y_i}})^2 (y_i-\bar{y_i})^2}},
\end{equation}

where $a$ and $b$ are the intercepts and the slope of the regression line, respectively.

\begin{figure}
  \centering
  \includegraphics[width=0.6\linewidth]{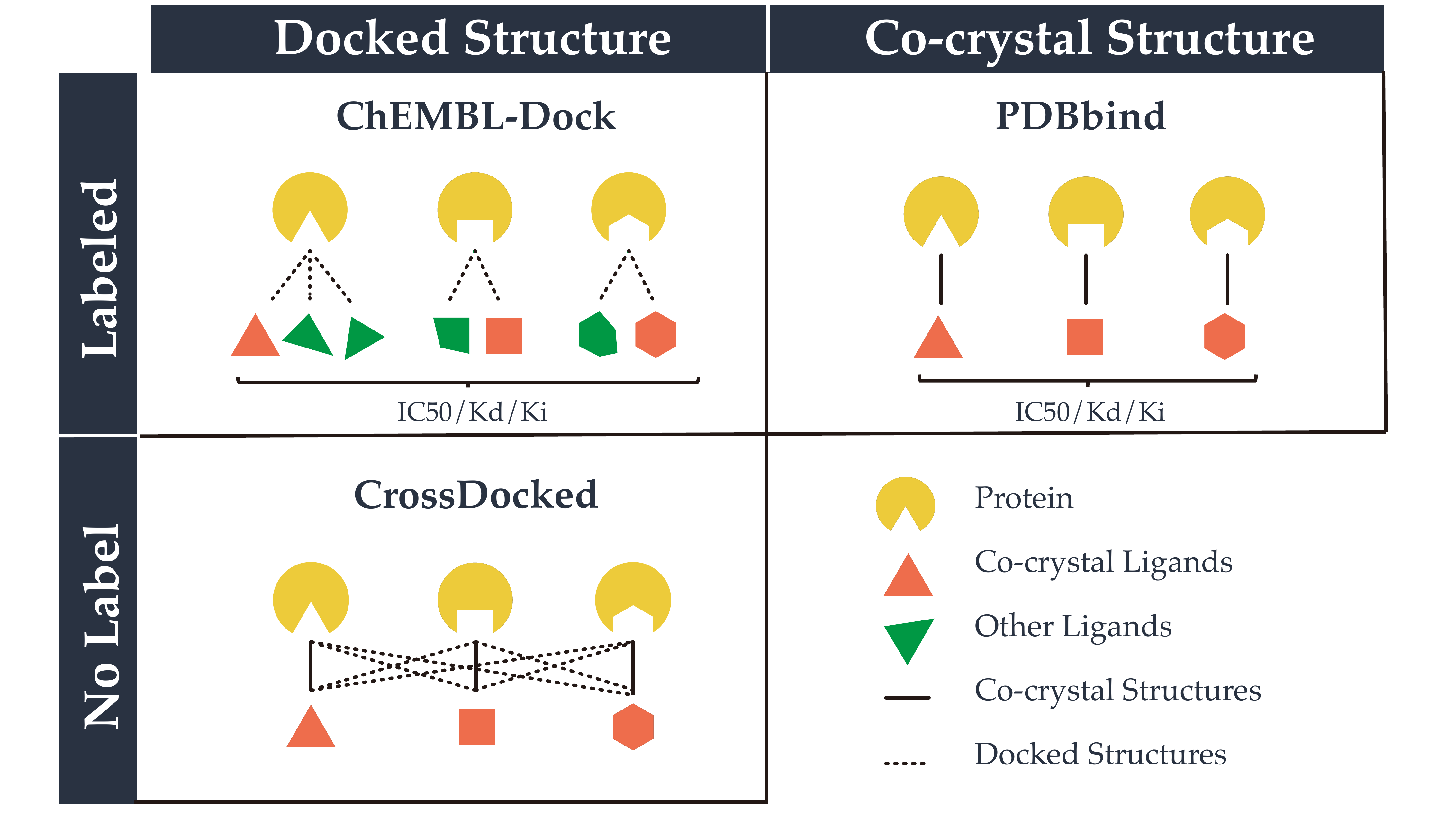}
  \caption{Comparison of \datasetname with PDBbind and CrossDocked on label and structure.}
  \label{fig:5_dataset_comparasion}
\end{figure}

\section{Curation workflow of \datasetname}
\label{Curation workflow of dataset}
The curation workflow of \datasetname is detailed in this section.

\vpara{Data Collection and Cleaning.}
We first clean up the database to select high-quality PLBA data, as the bioactivity database ChEMBL covers a broad range of data, including binding, functional, absorption, distribution, metabolism, and excretion (ADME) data.
The filtering criteria in ChEMBL we used are:
\begin{itemize}[leftmargin=*]
    \item STANDARD\_TYPE = `IC50'/`Ki'/`Kd' (other types of affinity, such as EC50, has limited bioassay data in ChEMBL);
    \item STANDARD\_RELATION = `=';
    \item STANDARD\_UNITS = `nM';
    \item ASSAY\_TYPE = `B' (meaning that the data is binding data);
    \item TARGET\_TYPE = `SINGLE PROTEIN';
    \item COMPONENT\_TYPE = `PROTEIN';
    \item MOLECULE\_TYPE = `Small molecule';
    \item BAO\_FORMAT = `BAO\_0000357' (meaning that in the assays, only results with single protein format were considered).
\end{itemize}
Besides the above filters, we further exclude assays with only one protein-ligand pair and assays with more than one affinity type.
We then prepare the 3D structures of proteins and  ligands, respectively.
The 3D structures of proteins are extracted from PDBbind using their Uniprot ID, while the 3D structures of  ligands are generated using RDKit.
We ignore all proteins not in PDBbind and ligands that RDkit fails to generate a conformation.
The final dataset comprises 313,224 protein-ligand pairs from 21,686 assays, containing a total of 231,948 IC50 protein-ligand pairs from 14,954 assays, 69,127 Ki protein-ligand pairs from 5,397 assays and 12,149 Kd protein-ligand pairs from 1,335 assays, respectively. 

\vpara{Molecular Docking.} Molecular docking software SMINA~\cite{Koes2013LessonsLI} is utilized to generate 3D protein-ligand complexes.
various docking strategies can be adapted to generate docking poses.
Given that the proteins included in our dataset are already present in the PDBbind database, we adopt a site-specific docking approach by specifying a search space.
The search space is a $22.5\mathrm{\mathring{A}} \times 22.5\mathrm{\mathring{A}} \times 22.5\mathrm{\mathring{A}}$ grid box centered on the ligand of the PDBbind complex which has the same proteins. For these data, we can accurately identify the binding sites of the protein-ligand complex from the 3D structure information provided in the PDBbind database. 
During the docking process, SMINA generates 9 candidate poses for each protein-ligand pair. Consequently, we obtain a total of 2,819,016 poses for the 313,224 protein-ligand pairs included in our dataset. In our study, we exclusively consider the top-ranked pose for further analysis and evaluation.

\section{Additional experimental results}
\label{Additional experimental results}
In this section, we conduct a comprehensive comparative analysis between MBP, TANKBind, and Transformer-M.
To ensure a fair evaluation, we base our comparison on the reported results from their papers.
\subsection{Comparision with Transformer-M}
We follow Transformer-M's methodology to train and test on the PDBbind2016 dataset and use the same dataset split as Transformer-M (see \url{https://openreview.net/forum?id=vZTp1oPV3PC&noteId=ZEa3K6qePg_ for more details}). Following Transformer-M, we employ the adversarial training method FLAG\cite{Kong2020RobustOA} during fine-tuning. We repeated the experiment five times.
Our comparative analysis reveals that MBP consistently outperforms Transformer-M across multiple evaluation metrics, including
RMSE, SD, and R.
Notably, MBP achieves these superior results while utilizing significantly fewer model parameters, with only about 1 million parameters compared to Transformer-M's 50 million parameters.
These findings demonstrate the remarkable effectiveness of MBP in the task of PLBA prediction, surpassing the performance of the powerful molecule pre-training method, Transformer-M.

\begin{table*}
\centering
\caption{Comparision with Transformer-M on Transformer-M's setting. The mean RMSE, MAE, SD, and R (std) over 5 repetitions are reported. The best results are highlighted in \textbf{bold}.}
\label{tbl:transformer-m}
\scalebox{1.0}{
\small
\begin{tabular}{c|cccc}
\toprule
\multirow{2}{*}{Method}& 
\multicolumn{4}{c}{PDBbind core set}\\
\cmidrule{2-5}
& RMSE$\downarrow$ & MAE$\downarrow$ & SD$\downarrow$ & R$\uparrow$ \\
\midrule
Transformer-M & 1.232 (0.013) & \textbf{0.940 (0.006)} & 1.207 (0.007) & 0.830 (0.011) \\ 
MBP & \textbf{1.208 (0.011)} & 0.943 (0.011) & \textbf{1.164 (0.017)} & \textbf{0.845 (0.005)} \\ 
\bottomrule
\end{tabular}
}
\end{table*}

\subsection{Comparision with TANKBind}
In this section, we present a comparison between MBP and the molecular docking method TANKBind in the task of PLBA prediction. Specifically, we train and test MBP on the PDBbind2020 dataset, utilizing the identical dataset split (time split strategy) as employed by TANKBind. To establish statistical robustness, we repeated the experiment five times, consistent with the methodology of TANKBind.
The obtained results, as summarized in Table S2, reveal a substantial advantage of MBP over TANKBind across various evaluation metrics, including RMSE, MAE, and R. Notably, MBP demonstrates a remarkable 6.6\% improvement in terms of the RMSE metric when compared to TANKBind.
These findings substantiate the competitiveness and effectiveness of MBP.

\begin{table*}
\centering
\caption{Comparision with TANKBind on TANKBind's setting. The mean RMSE, MAE, SD, and R (std) over 3 repetitions are reported. The best results are highlighted in \textbf{bold}.}
\label{tbl:tankbind}
\scalebox{1.0}{
\small
\begin{tabular}{c|cccc}
\toprule
\multirow{2}{*}{Method}& 
\multicolumn{4}{c}{Time split test set}\\
\cmidrule{2-5}
& RMSE$\downarrow$ & MAE$\downarrow$ & Spearman$\uparrow$ & R$\uparrow$ \\
\midrule
TANKBind & 1.346 (0.007) & 1.070 (0.019) & \textbf{0.703 (0.017)} & 0.726 (0.007) \\ 
MBP & \textbf{1.257 (0.002)} & \textbf{1.012 (0.016)} & 0.697 (0.008) & \textbf{0.736 (0.001)} \\ 
\bottomrule
\end{tabular}
}
\end{table*}


\begin{figure}[ht]
  \centering
  \includegraphics[width=0.8\linewidth]{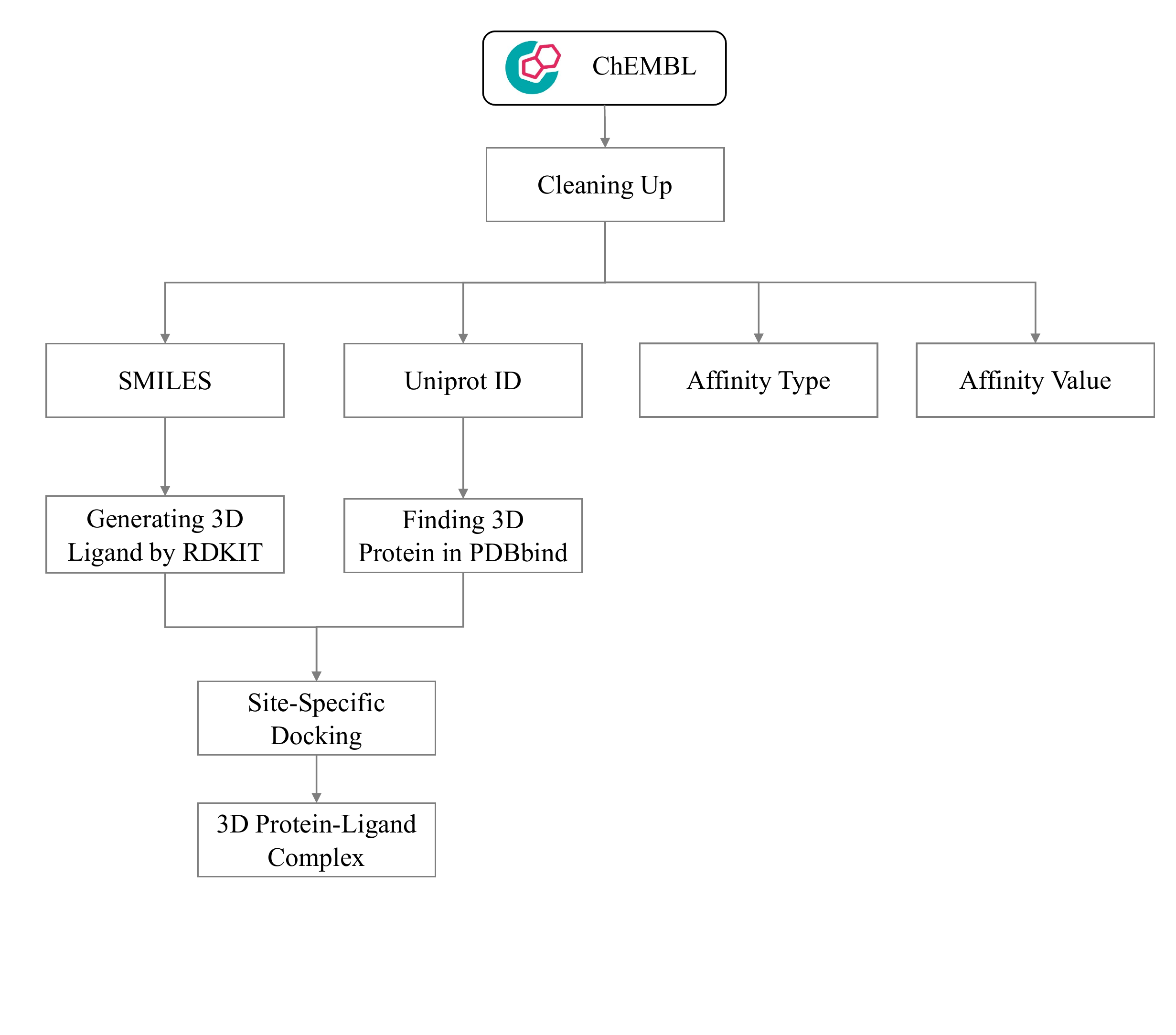} 
  \vspace{-0.5in}
  \caption{Construction process of \datasetname}
  \label{fig:appendix_dataset}
 \end{figure}
\end{document}